# The fundamental diagrams of elderly pedestrian flow in straight corridors under different densities

Xiangxia Ren[a], Jun Zhang[a*], Weiguo Song[a], Shuchao Cao[b]

[a]*State Key Laboratory of Fire Science, University of Science and Technology of China, Hefei 230027, China*

[b]*School of Automotive and Traffic Engineering, Jiangsu University, Zhenjiang 212013, China*

**Abstract:**

The worldwide population is aging and countries are facing ongoing challenges in meeting the transportation demand of the elderly. In this study, we investigate the movement characteristics of elders in the straight corridor and compare them with that of young adults. The free speeds of the elderly (about 1.28m/s) are obviously slower than that of the young people (about 1.4m/s) in the laboratory experiments. It is found that the fundamental diagram of the elderly shows the similar trend compared with that of young pedestrians. However, at the same densities the speeds of the elderly are always lower than that of the young pedestrians in the observed density range (< 3.0 m$^{-2}$). When the mean velocity calculated from pedestrian movement at low densities is considered, the two normalized fundamental diagrams agree quite well. The reasons for the differences are explored by analyzing the border distance, the nearest neighbors as well as the spatial distribution areas of the pedestrians. Our findings can be useful for the improvement of the pedestrian modelling and the design of pedestrian facilities that are much friendlier to the elderly.

**Keywords:** elderly pedestrian; straight corridors; fundamental diagram; free speed; spatial distribution;

## 1. Introduction

Research on the pedestrian dynamics is the key to realize safer transportation environments and effective evacuation in emergencies. Pedestrian experiments under well controlled but various conditions were carried out to study the movement of pedestrians [1-7]. The current research indicates that the building structure, the motivation of the crowd and the data collection methods have obvious influences on the fundamental diagram of the pedestrian movement [8-10]. Experiment of pedestrians walking in the corridors were carried out to study the relationship between movement characteristics and corridor conditions [1, 11, 12]. Based on the experimental data, some simulation models [13-15] were developed to study the characteristics of pedestrian dynamics deeply, especially some experimental conditions that cannot be realized in practice.

On the other hand, the aging of population is a social phenomenon around the world. Compared with young people, elders are vulnerable group and facing with higher risks in daily life especially

---

in the transportation due to their poor mobility, vision, hearing, strength and sensory ability. An evacuation experiment of moving downstairs conducted by Kuligowski [16] showed that the elderly people have lower velocities and longer evacuation time requirements. The elderly transportation demand is increasing day by day and it is of great significance to study the motion characteristics of the elderly population and to design more friendly facilities so as to improve the safety of the elderly. How to realize efficient evacuation and safer traffic of elders by considering their special physical conditions is a great challenge for public safety work. However, the studies on the movement behavior and characteristics of elders are still less.

Most of data in the field of pedestrian dynamics is on young and middle-aged people. The differences of the movement dynamic characteristics between the elders and the young people are still not clear. Observational experiments [17-20] were carried out in different places to observe the fundamental characteristics of elderly pedestrian and it is found that factors such as gender, disease, educational level, grip strength can affect the walking speed. The results of the field observation conducted by Gorrini et al. [21] in an urban crowded walkway show that in situation of irregular flows elderly pedestrians walked 40% slower than adults, but the influence of group and environment on the movement cannot be ruled out. Data from the single-file experiment in [22, 23] showed that there are significant differences in the fundamental diagrams between the elderly and mixed group and the ratio of the elderly in the group has different influence. Further, the curvature of trajectory is analyzed [24] and it is found that elders have larger step widths for they have a weaker ability to control the lateral movement. Where the step width was defined as lateral distance between two consecutive steps. However, this experiment only considered a single-file scenarios and lacks high-density data.

Spatial-temporal distribution is an important self-organization features of pedestrian. It is found that there are symptoms of strong correlations between positions of closely located pedestrians [25]. The age, gender, and mobility also play important roles in interpersonal distance [26]. Compared with young people, the elderly need farther international distance [27]. However, there is still a lack of quantitative description, and the specific expression of spatial distribution in the crowd also deserves our attention, especially for the elderly.

Some of the simulations take the age parameters into account [28-30]. Shimura et al. built a CA model according to the characteristics of the pedestrians' mobility when the elderly and the young are mixed to gain some basic phenomena of dynamics [31]. But at the moment, motion models for the elderly is still very limited due to the lack of basic empirical data. Research on the dynamics of the elderly needs to be supported by richer data and the effect of age on the movement characteristics of group is still lack of in-depth study.

Based on these considerations, we implemented an experiment to investigate the movement of elderly pedestrian in straight corridors under different density. The aim of our study is to compare the fundamental diagrams of elderly group with groups of different ages and analyze the influence of age on the crowd movement in the corridors. The structure of this paper is as follows. In section 2 we describe the setup of the experiment and the trajectories in different scenes. We analyze the fundamental diagrams of the elderly movement in straight corridor and compare the results with previous studies on young adults in section 3. Finally, section 4 summarizes the paper and makes a conclusion.

## 2. Setup of experiment

The experiments were carried out on March 2018 in Hefei, China. We recruited 73 volunteers from a senior center in Hefei. Fig.1 shows the distribution of age and gender of the participants. They are between 52 to 81 years old with the mean age of 69.7±7 years old. Their heights range from 150cm to 175cm, with an average of 163cm. The ratio of male and female is about 1:2.5 (21 males and 52 females). Even though they have no physical problem for normal movement, we did not ask them to participate in each run of the experiment by taking account of their demands for rest.

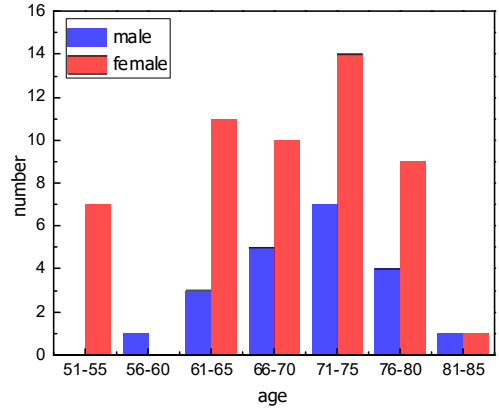

Fig.1. The distribution of age and gender of the pedestrians in the experiments.

Fig.2 shows the illustration and a screenshot of the experimental scenario. We studied the movement of elderly in straight corridor under controlled conditions. The length and the width of the channel is 10m and 1.8m respectively. In order to regulate and control the density in the corridor, nine runs were designed by changing the sizes of the entrances and exits of the corridors. The detailed information on each scenario setup can be found in Table 1. Before the formal experiment, we asked each participant to walk through the corridor once to measure their free movement speed afterward. When the experiment started, they were asked to walk through the corridor in normal way from the waiting area 2m away from the entrance. Note that few participants asked to have a rest after some runs due to their physical abilities. This is the reason why the number of pedestrians in different runs did not remain the same. At the end of the experiment, a short questionnaire was used to collect their personal information including age, sex, height, weight and so on.

We used two digital cameras mounted on the roof of a building about 10m high to record the process of these experiments. Each participant was asked to wear a red or blue hat for easily recognition from video recordings (see Fig.2). The software *PeTrack* [32] was used to extract the trajectories automatically and the average height of 163cm is used for data transformation from pixel coordinates to physical coordinates.

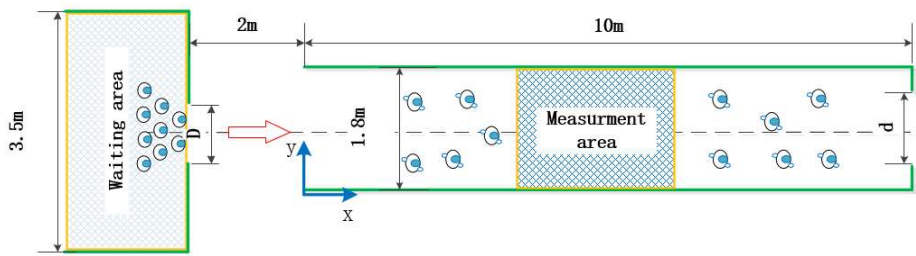

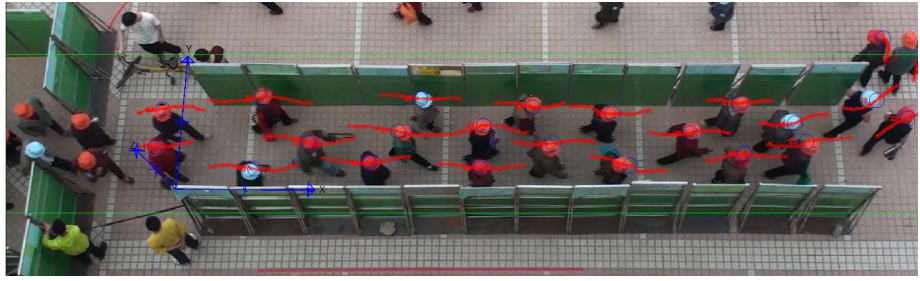

Fig.2. The sketch (above) and a screenshot (below) of the scenario, we changed the width of the inlet D and the outlet d to form different densities in the straight corridor.

Table 1. Controlled conditions and number of participants in every experiment.

| Index | Conditions | Number of pedestrians |
| --- | --- | --- |
| C-050-180-180 | D=0.5m, d=1.8m | 64 |
| C-070-180-180 | D=0.7m, d=1.8m | 63 |
| C-090-180-180 | D=0.9m, d=1.8m | 73 |
| C-110-180-180 | D=1.1m, d=1.8m | 55 |
| C-180-180-180 | D=1.8m, d=1.8m | 62 |
| C-180-180-160 | D=1.8m, d=1.6m | 67 |
| C-180-180-130 | D=1.8m, d=1.3m | 67 |
| C-180-180-100 | D=1.8m, d=1.0m | 64 |
| C-180-180-070 | D=1.8m, d=0.7m | 59 |

Fig.3 shows the pedestrian trajectories obtained from the video recordings with instantaneous speed in different scenarios. In the process of the movement, the pedestrians formed three lanes and converged around the outlet. With the decrease of the outlet width more fluctuations appear in the trajectories, since pedestrians exhibit a physical swing during the process of waiting and finding the path under high densities. At the same time, the velocity of the crowd is decreasing, especially in the area near the exit. There is a sharp drop in velocity at the inside of the corner. Itcan be found from these graphs that the narrower the exit, the greater the impact on pedestrian velocity. These trajectories also indicate that a part of people have a higher velocity for they get through the entrance as quickly as possible in order to avoid congestion. By observing the experiment video, we found that these pedestrians are those with better physical conditions and higher participation enthusiasm in the experiment. Based on these trajectories, movement characteristics of the elders such as density, velocity and flow at any time and position can be calculated.

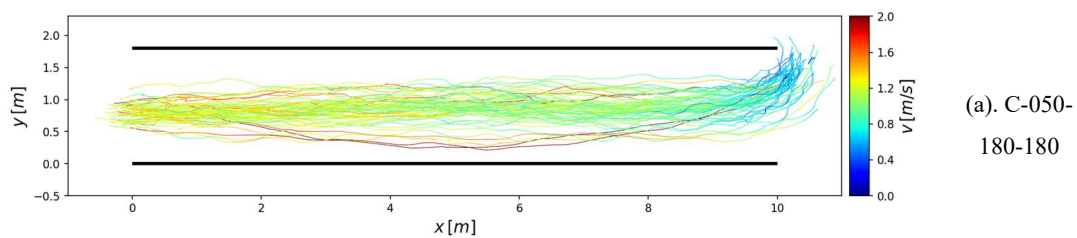

(a). C-050-180-180

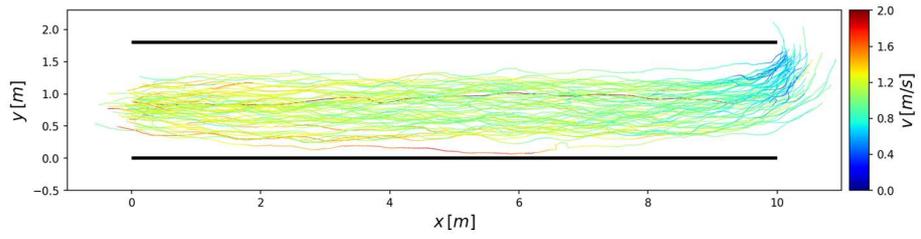

(b). C-070-180-180

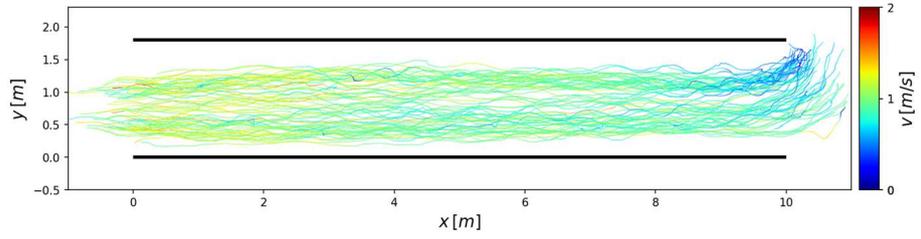

(c). C-090-180-180

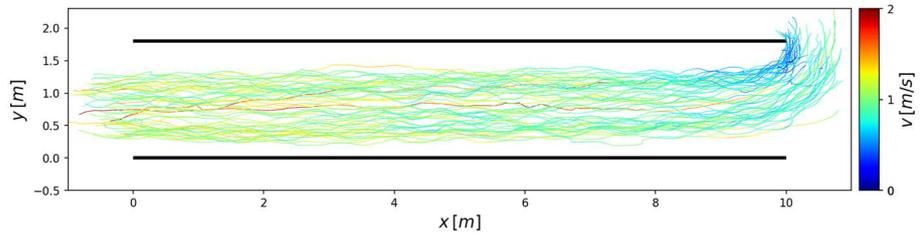

(d). C-110-180-180

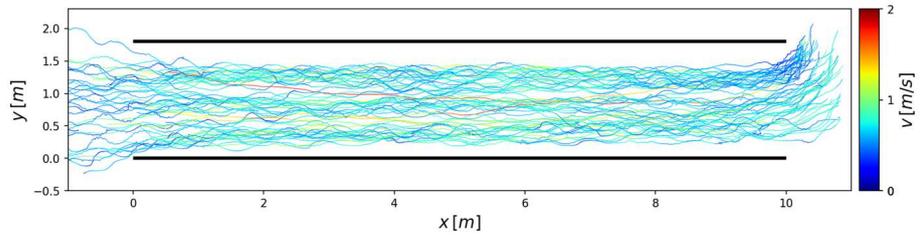

(e). C-180-180-180

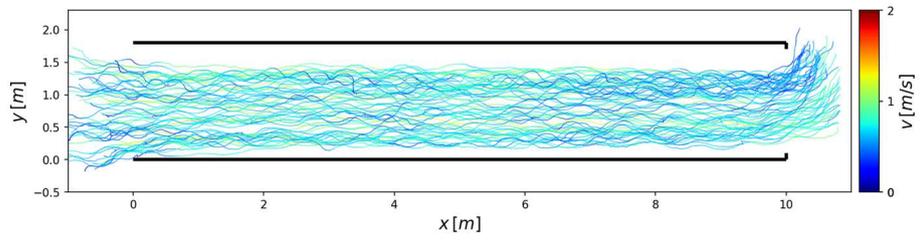

(f). C-180-180-160

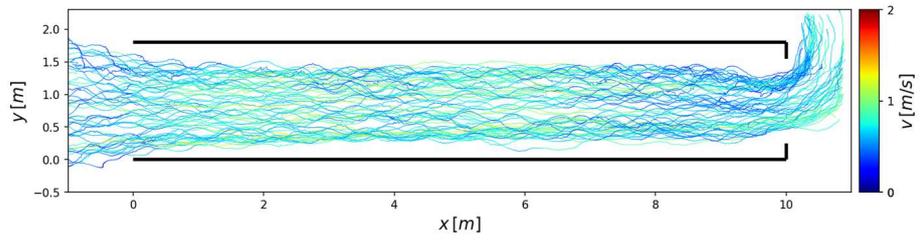

(g). C-180-180-130

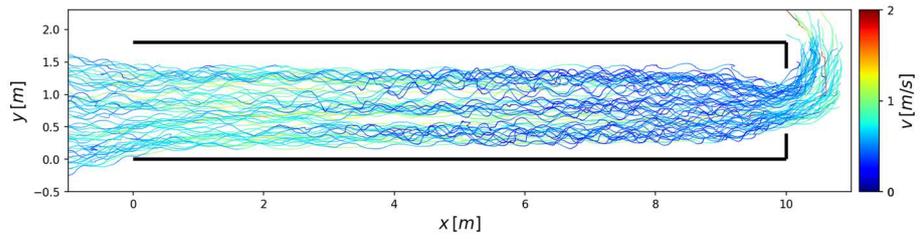

(h). C-180-180-100

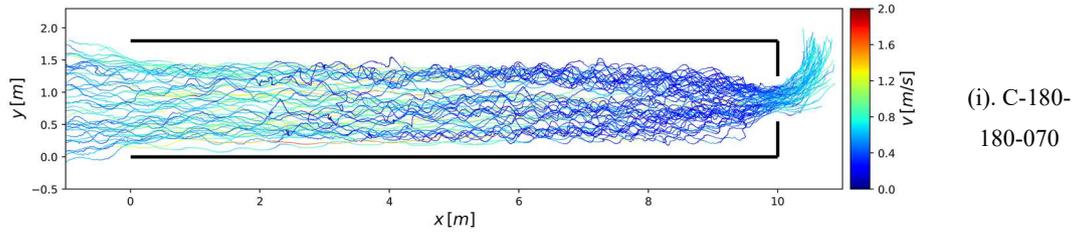

(i). C-180-180-070

Fig.3. Trajectories with instantaneous velocity in different scenarios.

## 3. Analysis and results

### 3.1 Fundamental diagram

Firstly, we study the fundamental diagrams of unidirectional flow of elders in the corridor using the Voronoi based method introduced in [10]. In order to investigate the effect of different measurement areas on the results, densities and velocities in five different areas with the size of 2m×1.8m in the corridor along the movement direction are measured. The detailed locations of the measurement areas are listed in Table 2. To determine the fundamental diagram, only data from stationary flows are used which are selected by analyzing the time series of density and velocity (see appendix). To compare with previous data, the same time interval $\Delta t$ =0.4s (corresponding to 10 frames) is selected to calculate the instantaneous velocity of each elder. In Fig.4 we show only one point per 10 frames to limit the number of sample points and to guarantee their independence. We can see that the fundamental diagrams obtained from different measurement areas agree well except for the difference in data range. When the measurement area is closer to the exit, higher and more continuous densities can be obtained, which makes the fundamental diagram smoother. The shape of the fundamental diagram is similar with that in previous studies on young pedestrian movement [10]. Note that these young people are German, it is not sure how will the cultural differences influence the results. With the increasing density, the velocity of the crowd declines continuously while the specific flow $J_s$ increases first and then decreases. The critical density where the specific flow reaches the maximum is about 1.5 m$^{-2}$, which is smaller than that of young pedestrian flow. However, the minimum width of exit (about 1.3 m see appendix) for reaching the maximal $J_s$ is consistent with Hankin's findings [33]. He found that above a certain minimum of about 4 ft (about 1.22 m) the maximum flow in subways is directly proportional to the width of the corridor.

Table. 2. The detail of the measurement areas that were selected to measure the fundamental diagrams.

| Index | Range of x(m) | Range of y(m) |
| --- | --- | --- |
| a | 2-4 | 0-1.8 |
| b | 3-5 | 0-1.8 |
| c | 4-6 | 0-1.8 |
| d | 5-7 | 0-1.8 |
| e | 6-8 | 0-1.8 |

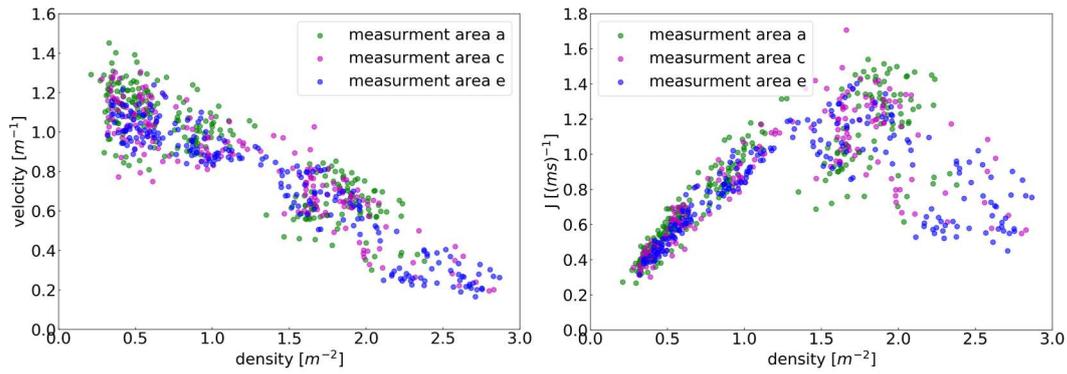

(a). Relationship between velocity and density.   (b). Relationship between specific flow and density.

Fig.4. Fundamental diagrams obtained from three different measurement areas.

Since the same dimension of the experimental corridor was set, it is possible to compare the difference of the fundamental diagrams of unidirectional flow for the elders and German young people in [10]. The same size of the measurement areas ($\Delta x \in$ [4m, 6m] for the elders, $\Delta x \in$ [3m, 5m] for the young) in the middle of the channels are selected in both experiment and the same data analysis process are used here. From Fig. 5, we can see that the shapes of fundamental diagrams for both kinds of population are similar from the scatter diagrams. The difference is that the data points for the elderly are always below that of the young. In other words, at the same density, the velocity and specific flow rate (J) of the elderly is lower than that of young. The observed difference confirms the truth that the elderly have lower athletic ability than young pedestrians do. Under the same situation, the capacity of accessing a facility is lower for the elderly.

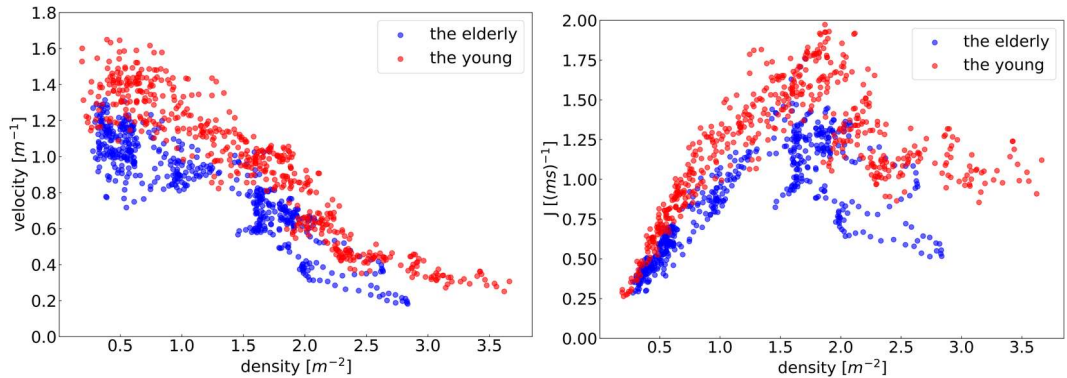

Fig.5. Comparison of the fundamental diagrams of unidirectional flow of the elderly and the young.

What's more, the velocity difference becomes smaller with the increase of the density in the density-velocity diagram. It implies that the influence of mobility on pedestrian flow decreases when the density becomes higher. When the density is relatively small, there is enough space for pedestrians to move freely. Pedestrian movement depends more on the individual mobility and the difference between the two groups is relatively large. As the density increases, the distance among pedestrians become short and interactions among them get stronger.

To understand the mechanism of the different phenomena, we analyzed the movement characteristics from the aspects of free speed, boundary distance and spatial distribution in the following sections.

## 3.2 Statistic analysis of free speed

The velocities of each pedestrian over the entire length of the channel were calculated, we found that the differences between the pedestrians are relatively large (Fig.6). In consideration of the small sample size, different measurement areas were adopted to statistics of the pedestrian free speed. The details of the measurement area is shown in Table.3. In order to increase the sample capacity, we computed the velocities of every pedestrian in 0.5m long area (3m-3.5m, 6m-6.5m), 1m long area (5m-6m), 2m long area (0m-2m, 2m-4m, 4m-6m, 6m-8m), 3m long area (0m-3m, 3m-6m, 6m-9m), 4m long area (0m-4m, 4m-8m), and 8m length range in the channel (0m-8m). Half participants were selected to show the effects of regional changes on free velocity. It can be seen from Fig.7 (a) that the velocity value of each measurement area does not fluctuate widely. In this way, we can get the numerical distribution of the free velocity of pedestrian in different length range and different passage area (Fig.7 (b)). The average free velocity of female is 1.30±0.18 m/s, while it is 1.22±0.21 m/s for male. We got p=0.000 through T-test which represents a significant difference between this two group of velocities. The speed of the elderly is significantly lower than that of young people (1.4 m/s). From the scatter plots and distribution maps, we can see that most elders have a speed of less than 1.4 m/s. Moreover, their speeds are roughly in line with the normal distribution.

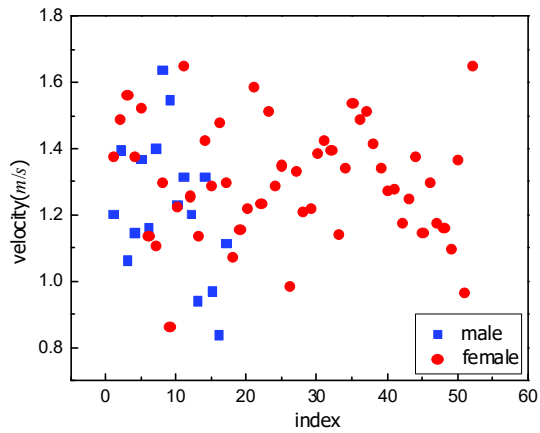

Fig.6. The free velocity of each pedestrian in the entire corridor, the squares represent velocity of male and the full circles represent that of female.

Table. 3. The details of the measurement areas selected to calculate the free velocity of each pedestrian.

| Length of the measurement area(m) | Specific X-axis range | Index |
| --- | --- | --- |
| 0.5 | 3-3.5 m | a |
|  | 6-6.5 m | b |
| 1 | 5-6 m | c |
| 2 | 0-2 m | d |
|  | 2-4 m | e |
|  | 4-6 m | f |
|  | 6-8 m | g |
| 3 | 0-3 m | h |
|  | 3-6 m | i |
|  | 6-9 m | j |
| 4 | 0-4 m | k |

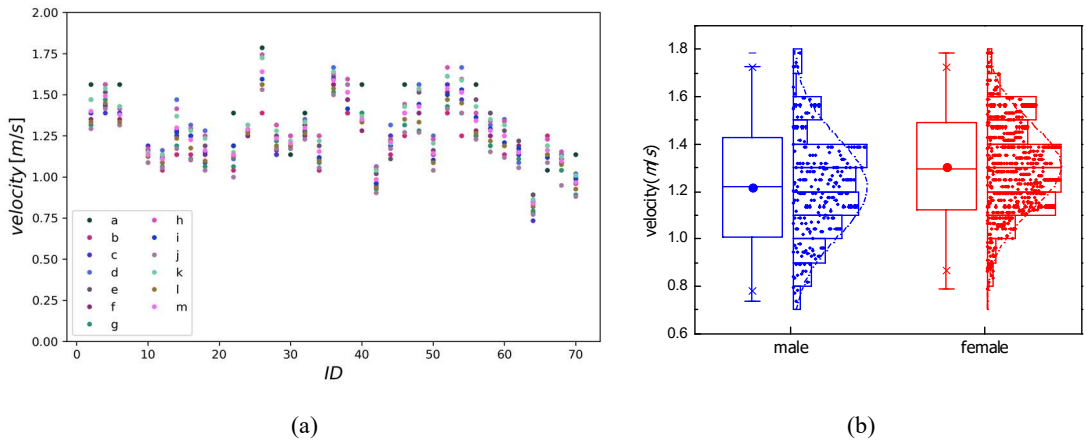

(a)          (b)

Fig.7. Free velocity of each participant in different measurement area.

Overall, elders have a lower free speed than young. The female elders are slightly faster than male. For the most important thing is that women are more active than men in our experiment area. In addition, most of the women have peers, they form a weak competition between them. On the other hand, this can be attributed to women's slower aging than men's, so the inhomogeneity of older women is more pronounced than older men. Due to the small sample size and unbalanced sex ratio, the results only represents for this experiment and general applicability of the results remains to be verified.

To explore the impact of free velocity on the movement characteristics of elders and young further, we normalize the velocity based on the free velocity and replot the fundamental diagrams in two ways as shown in Fig.8. Firstly, we use the directly measured free velocities $V_{f1}$ before experiment: 1.28m/s for the elderly group in our experiment and 1.4 m/s for the young adults. As shown in the Fig.8 (a), the difference between the two sets of data is significantly reduced. Nevertheless, the data point for the elderly is still below the young. Furthermore, we suppose that the pedestrians move in free speed $V_{f2}$ when the density is less than 0.7m$^{-2}$, where the velocity seems independent on density. Consequently, the mean value of the free velocity for the elderly and young adult experiment are 1.07m/s and 1.37m/s respectively. Therefore, the expected velocity relation coefficient between the two groups of pedestrians is 1.28. The normalized fundamental diagrams shown in Fig.8 (b) agree very well excepting when the density is higher than 2.5 m$^{-2}$ where the small amount of data caused the error. The differences of the two normalizations can be analyzed from the following aspects. The mean velocities $V_{f2}$ in low density takes the weak interactions among people into account, which reflects the free movement of the crowd better than individual movement. While the free velocities $V_{f1}$ were measured by asking the pedestrian to walk in the corridor individually at the beginning of the experiment. When one was asked to walk alone, he or she will be in a relative excitement state to maintain a better personal image. Besides, the physical strength is another point that need to be mentioned. As the experiment went on, the fatigue will easy increase and the enthusiasm decreases especially for the elderly.

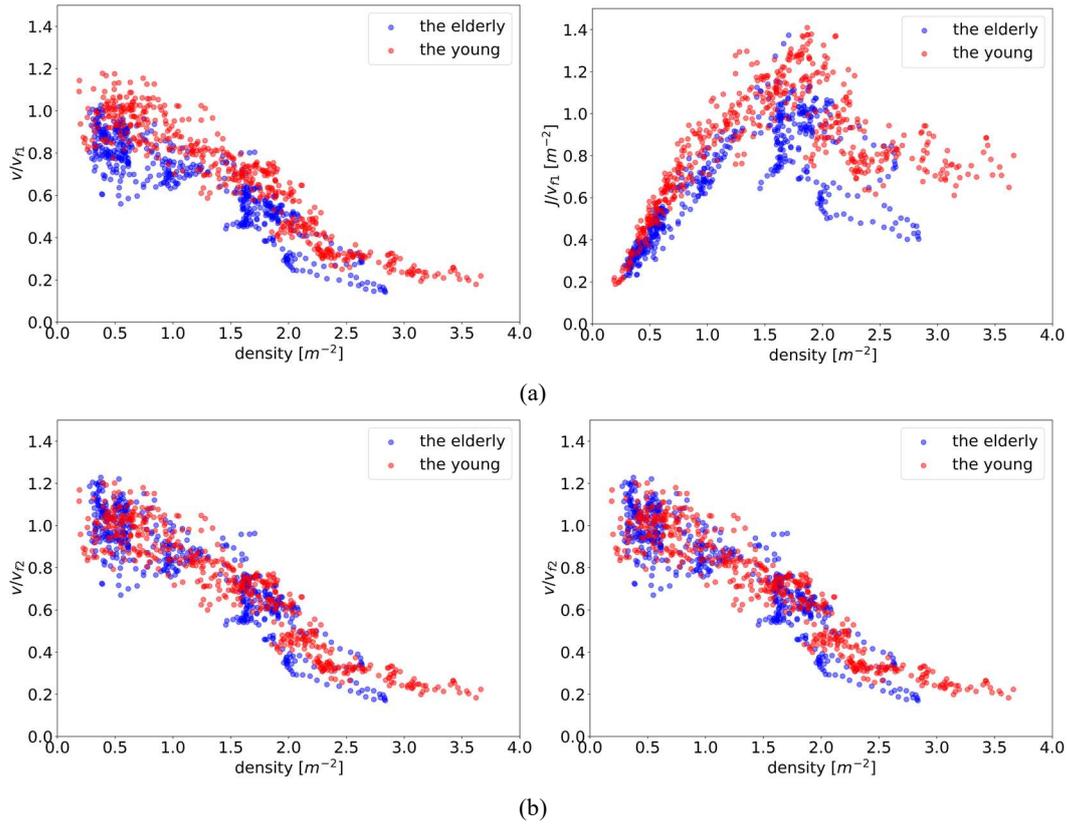

Fig.8. Normalized fundamental diagrams based on the free velocities obtained in two different ways, (a) directly measured free velocity from individual movement, (b) the mean velocity for pedestrian movement under low densities.

**3.3 Distance from the boundary**

People unconsciously keep certain distance to walls when walking. Compared with the trajectories in [34] whose participants were 25±5.7 years old, the elders in this experiment are farther away from the boundary. In the video, it can be observed that the participants deliberately keep distance from both walls to avoid collisions. In order to quantify it accurately, the distances for each pedestrian to its nearest wall are calculated under different global densities. It is noticed that we assume the average shoulder width of the pedestrians is 0.4m here. Then the distance $d_{pw}$ from the shoulder of a pedestrian at the border of the crowd to the wall is calculated. To make comparison, we calculate the time series of the mean distances to the wall under four different densities (around 1.6 $m^{-2}$, 1.8 $m^{-2}$, 2.0 $m^{-2}$ and 2.2 $m^{-2}$) from the experiments with elderly and young people respectively. In order to quantify the size relationship between the elders and young people further, both the mean distance to the wall and the variance were plotted in Fig.9. And Table.4 shows the specific value of the different densities and boundary distance for the two groups under the four situations. Obviously, the distance $d_{pw}$ for the elderly are always larger than that for the young nearly for all the four different densities. Further, $d_{pw}$ can be regarded as a constant especially for the elderly. However, the standard deviation of $d_{pw}$ is relatively large under each density due to the swing phenomenon.

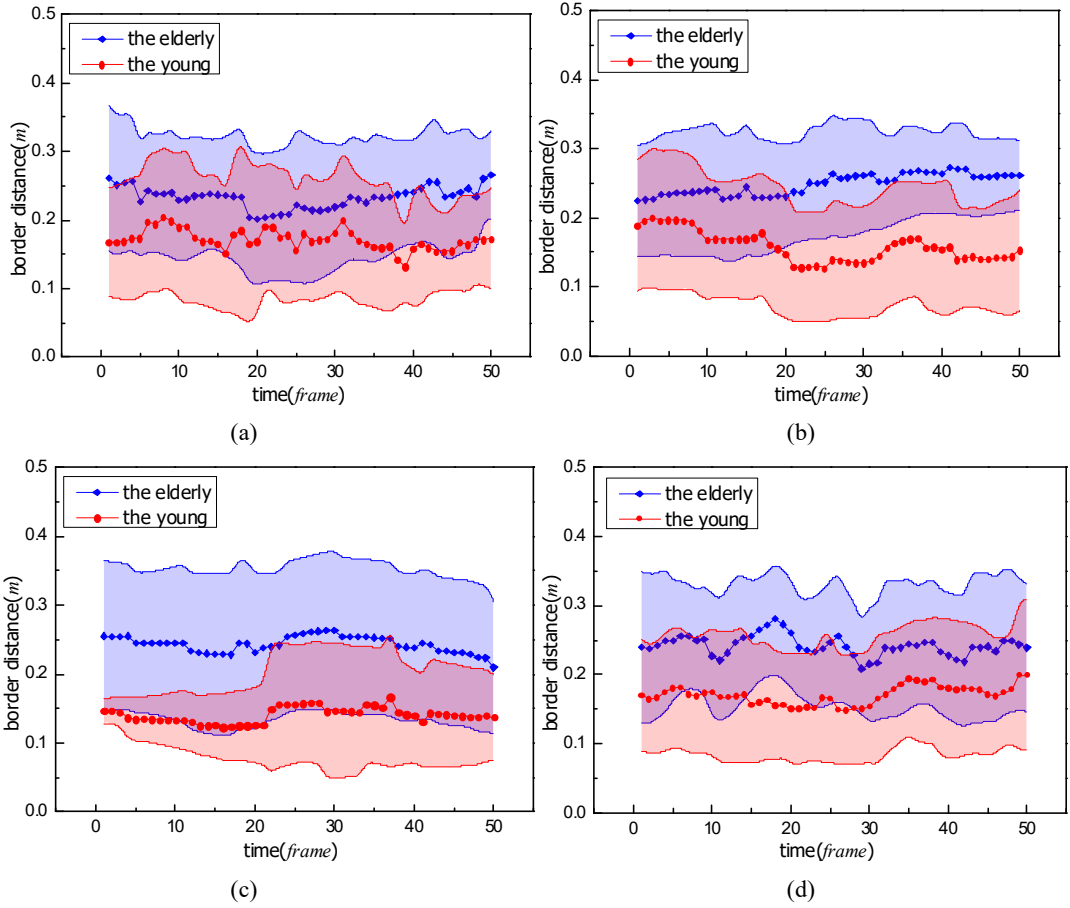

Fig.9. Distances to the wall of the elders compared with that of the young in same densities, the a, b, c, d corresponding to four densities (around 1.6 m$^{-2}$, 1.8 m$^{-2}$, 2.0 m$^{-2}$ and 2.2 m$^{-2}$).

Table 4. The mean border distance d$_{pw}$ under different densities.

| Index | Density of the elderly | Distance of the elderly | Density of the young | Distance of the young |
| --- | --- | --- | --- | --- |
| a | 1.64 m$^{-2}$ | 0.23±0.09m | 1.68 m$^{-2}$ | 0.17±0.09 m |
| b | 1.83 m$^{-2}$ | 0.25±0.08m | 1.79 m$^{-2}$ | 0.16±0.09 m |
| c | 1.99 m$^{-2}$ | 0.24±0.11m | 1.97 m$^{-2}$ | 0.14±0.06m |
| d | 2.21 m$^{-2}$ | 0.24±0.09m | 2.22 m$^{-2}$ | 0.17±0.08m |

We conjecture that the larger d$_{pw}$ of the elderly mainly results from the following three aspects including physical conditions, environmental factors, and psychological effects. Firstly, the body sizes of the elderly in our experiment are thinner than that of the young adults in [34]. In the corridor with a given width, pedestrian with larger body occupies more space that can lead the neighbors to be close to the wall. Unfortunately, the participants' body size were not measured in both experiments and we could not quantify the effect of body size on d$_{pw}$ in this study. Secondly, it can be attributed to the adaptability of different groups to the environment and the different mental mechanism. The elderly are more alert to the unfamiliar circumstances compared to young people. They confirmed the purpose and safety of the experiment many times before the experiment, although we did a full explanation and informed them of the security. They need to be ready for any potential dangers all the time because of the loss of athletic ability so they prefer to stay much closer to the other familiar pedestrians and away from the walls for the elderly population. Contrarily, young people are more independently and displays a different behavior. These reasons will not only cause the different $d_{pw}$ between the two groups, but also cause the difference in the spatial distribution of the population.

**3.4 Spatial distribution distance**

In this section, we analyze the spatial distribution distance of the two groups of people. The pedestrians influence each other during the movement, which is related to the spatial distribution distance between pedestrians. Therefore, it is value to pay attention to the concrete expression of the spatial distribution between pedestrians during the walking movement and the difference between elder people and young people. We analyzed the distances between individuals to their neighbors around from the trajectories.

One pedestrian has to be selected as the reference to calculate the distance and relative position of other pedestrians in the corridor. We have analyzed before that the boundary of both sides will affect the spatial distribution of pedestrians. In view of this, we select the pedestrians in the center of the crowd as the reference pedestrians. Considering that, the movement of the pedestrians is greatly affected by the pedestrians in front and both sides, distances and locations of pedestrian distributed at angles from 0° to 180° were analyzed. Here 0° represents the right side of a pedestrian and 90° represents the movement direction along the corridor. In the same way, we can calculate the nearest neighbors for all the pedestrians at any time $t_i$ under the relatively stable periods during which the global density of the crowd had no significant change. From Fig.10 and Fig.11, we can see that the spatial distribution of the first nearest neighbors for the elderly is similar to a half ellipse, whereas it is open for the young in the movement direction. It means that the nearest neighbors directly in front of the elderly is greater than that of the young. From the distribution histograms, it can be seen that the mean distance between the elderly is smaller than that of the young. We got p=0.000 in T-test which is shown that elders have smaller distances than the young people do.

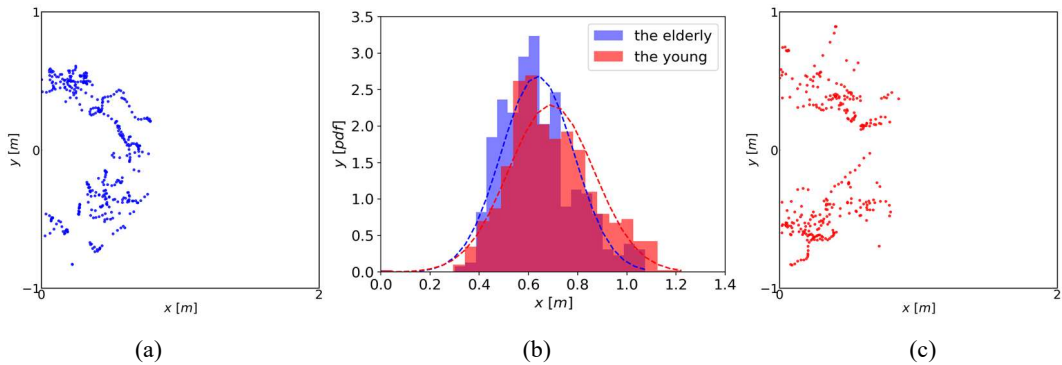

(a)        (b)        (c)

Fig.10 The distance and position distribution of the nearest neighbors for the density of 1.6 m$^{-2}$. (a) The nearest neighbors of the elderly. (b) Distribution of the nearest distance. (c) The nearest neighbors of the young.

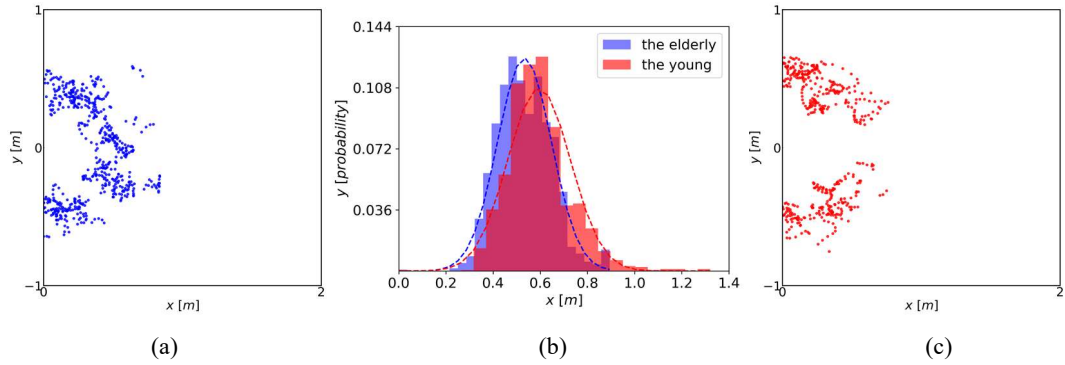

(a) (b) (c)

Fig.11. The distance and position distribution of the nearest neighbors for the density of 2.2 m$^{-2}$. (a) The nearest neighbors of the elderly. (b) Distribution of the nearest distance. (c) The nearest neighbors of the young.

If we suppose a pedestrian occupies a circle with the diameter of 0.4 m. The gap between two pedestrians $d_{pp}$ can be approximated as the distance between the head centers subtracting 0.4 m. In Fig.12 we compare the $d_{pp}$ and $d_{pw}$ between the elderly and the young under two different densities (1.6 m$^{-2}$ and 2.2 m$^{-2}$). In the case of low density, there is no significant difference between $d_{pp}$ and $d_{pw}$ of the elderly, whereas the $d_{pp}$ is obviously larger than $d_{pw}$ for the young people. The young pedestrians preferred to be close to the walls other than other pedestrians. However, with the increasing density the $d_{pp}$ decreases significantly while the $d_{pw}$ remains unchanged for both the young and the elders. What's more, there is no significant difference between $d_{pp}$ and $d_{pw}$ for the young people, but the $d_{pp}$ is obviously smaller than $d_{pw}$ for the elders. The repulsive force of border is greater for the elderly than for the young people.

The reasons are analyzed from the following three aspects. First, the distrust of the environment of the elderly leads to greater vigilance causing the boundary exclusion effect more obvious. Secondly, the familiarity among the participants can be a potential reason. The elderly volunteers were recruited from a senior center. They were relatively familiar with each other than the young pedestrians in the other experiment, which led to greater repulsive effect among the young people. Third, the effects of body size cannot be ignored, though we cannot quantify them yet. The larger body size of the young people from Germany may result in a bigger $d_{pp}$. However, this need to be checked in the future.

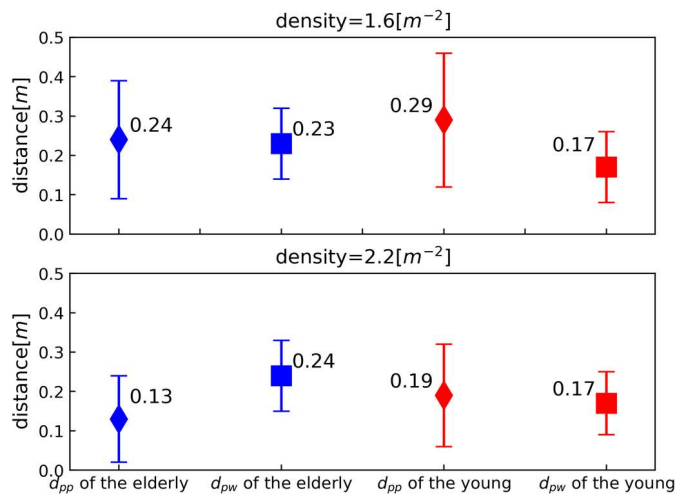

Fig.12 Comparison of $d_{pp}$ and $d_{pw}$ between the elderly and the young for density=1.6 m$^{-2}$ (above) and density=2.2 m$^{-2}$ (below).

## 3.5 Spatial distribution area

Both the nearest neighbor and others around affect the movement velocities of pedestrians. In the view of this, we search for the top six nearest neighbors of the reference pedestrians by sorting the distance between pedestrians and calculate the area of the polygons these six nearest neighbors make. The schematic diagram of the polygon is shown in Fig.13. The same method is used to calculate the polygon area composed by neighbors around the young adults. The specific values are presented in Table.5. It can be seen that the spatial areas of both groups decrease with the increasing density. Furthermore, the areas for the elders are always smaller than that of the young adults under the same global density (details can be seen from Fig. 14). This again proves that the elderly people preferred to stay closer each other and away from the walls, which leads to relatively smaller $d_{pp}$ and larger $d_{pw}$ correspondingly. One possible reason for this phenomenon is that they are more familiar with each other than the young adults are.

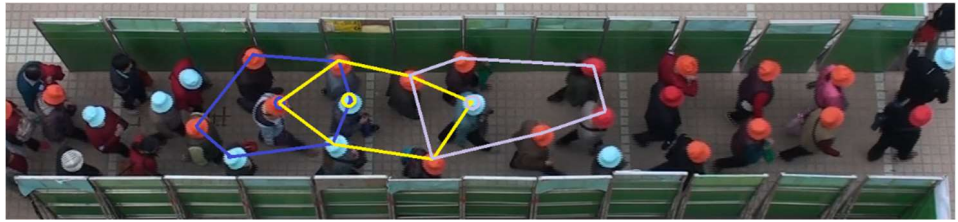

Fig.13. The polygon consisting of 6 nearest neighbors of any pedestrian

Table 5. The polygon area values in different conditions

| Index | Density of the elderly | Area of the elderly | Density of the young | Area of the young |
|---|---|---|---|---|
| a. | 1.64 m$^{-2}$ | 1.11±0.25 m$^2$ | 1.68 m$^{-2}$ | 1.43±0.29 m$^2$ |
| b. | 1.83 m$^{-2}$ | 1.08±0.24 m$^2$ | 1.79 m$^{-2}$ | 1.39±0.29 m$^2$ |
| c. | 1.99 m$^{-2}$ | 0.88±0.23 m$^2$ | 1.97 m$^{-2}$ | 1.19±0.22 m$^2$ |
| d. | 2.21 m$^{-2}$ | 0.81±0.18 m$^2$ | 2.22 m$^{-2}$ | 1.12±0.21 m$^2$ |

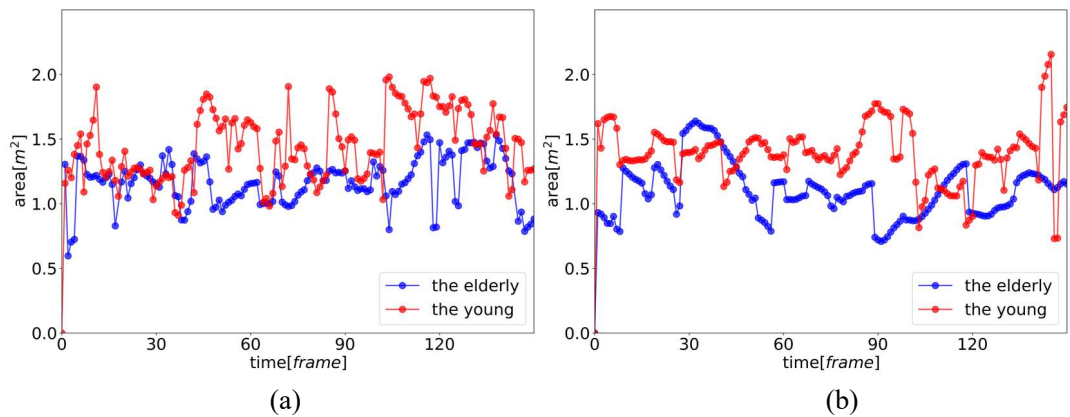

(a)　　　　　　　　　　　　　　(b)

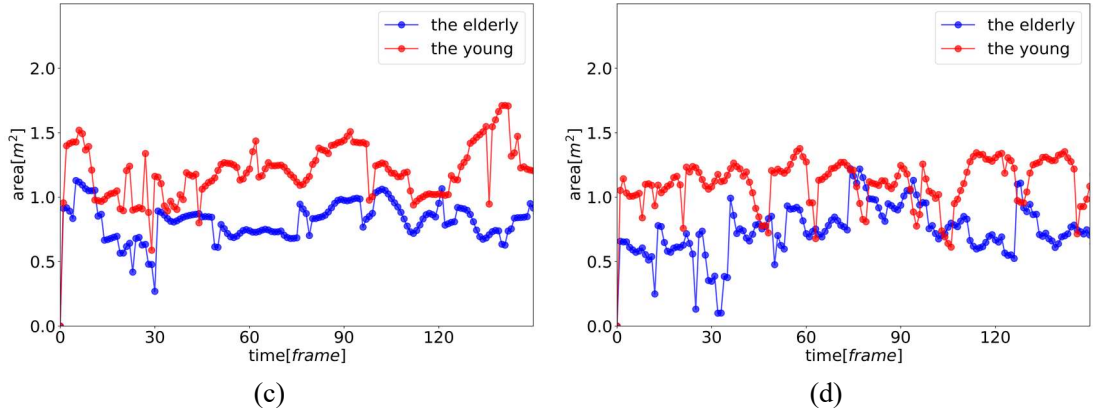

(c)                  (d)

Fig.14. Comparison of the time series of the spatial areas between the elders and the young people under the same densities, the a, b, c, d corresponding to densities (around 1.6 $m^{-2}$, 1.8 $m^{-2}$, 2.0 $m^{-2}$ and 2.2 $m^{-2}$).

Besides, we compare the spatial areas of the two group under the same velocities (for example, around 0.35m/s and 0.60m/s). Under these two situations, the densities in the corridor are relatively high and pedestrians distributed in three rows, which is fine to calculate the spatial area compared to low densities. As shown in Fig.15, interestingly there are no obvious difference between the elderly and the young for the both velocities. This seems to imply that the movement under high densities are less influenced by the movement ability of pedestrian. The speed can be constant if the spatial area for a pedestrian is the same regardless of old or young ones.

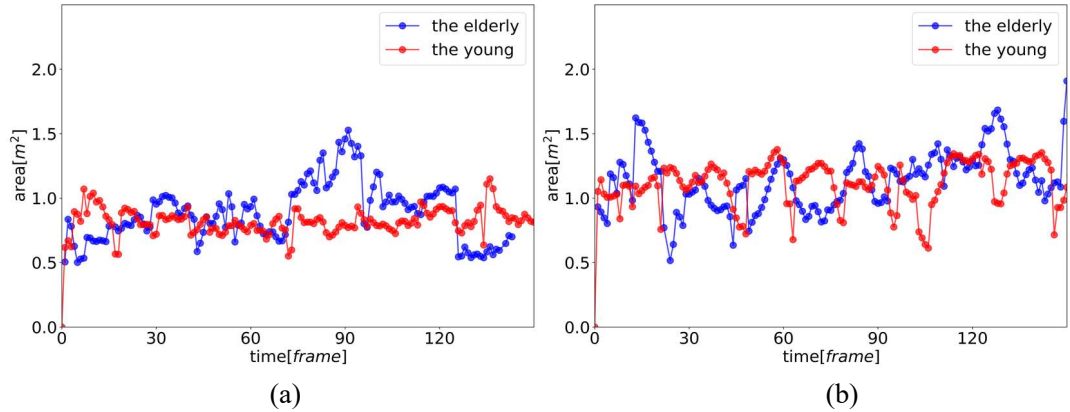

(a)                  (b)

Fig.15. Comparison of the time series of the spatial area between the elders and young people under the similar velocities. (a) 0.36m/s for elders and 0.34m/s for the young; (b) 0.57m/s for elders and 0.59m/s for the young.

## 4. Summary

In this study, walking experiment of elderly population in a straight corridor were performed under controlled conditions. In total, 70 elderly people over 60 years old participated in the experiment. We achieve different densities in the channel by changing the width of the entrance and exit and the experiment included nine different test scenarios. Compared with the young adults, the elderly show small steps, low pace and passive wait. Based on the pedestrian trajectories, the fundamental diagrams of the elderly are studied. It is found that the closer the measurement area is to the exit, the better the continuity of the fundamental diagram due to more data points. The basic shape of the fundamental diagram for the unidirectional flow of elderly is consistent with that in the previous researches. However, the speed and flow for the elderly are always smaller than the young adults under the same density situation are. This proves that elders in the same access facilities have lower capacity than young. We normalized the fundamental diagrams by considering the free velocity obtained in two different ways for the two groups. When the mean velocity calculated from pedestrian movement at very low densities is used, the two fundamental diagrams agrees well under the observed density range.

    Besides, in our experiment the average speed of the elderly is 1.218m/s for men and 1.304m/s for women, which is significantly smaller than that of the young adults (1.4 m/s). It is found that the elderly keep a larger distance to the walls than the young adults. We selected the pedestrian in the middle of the corridor at the steady state to calculate the distances and angles to his neighbors. Like the young, the nearest neighbors in the front of the elderly is oval-shaped. However, the spatial distance of the elderly is smaller than that of the young for 1.8m wide corridor. Furthermore, the area of the hexagon made up of the nearest six neighbors is calculated and compared. In the case of the same density, elder people have smaller areas than young people do, but there is no significant difference at the same speed. Elders gathering much closer than the young do during walking, which causes a greater mutual influence and further affects the velocity. We believe that differences in mobility lead to differences in spatial distribution between the two groups, which in turn reduces the speed and flow of the elderly population. The movement characteristics of the elderly obtained in the analysis can be useful for the design and construction of pedestrian facilities that are friendlier to the elders in the future. It is worth mentioning that the data used to compare with the elderly are from German students. The cultural difference between German and Chinese is not considered in this study and may also play a part on the results, which will be improved in the future work.

**Acknowledgement:**

The authors acknowledge the foundation support from the National Natural Science Foundation of China (Grant No. 71704168), from Anhui Provincial Natural Science Foundation (Grant No.1808 085MG217) and the Fundamental Research Funds for the Central Universities (Grant No. WK2320000040).

**References:**

[1] Hoogendoorn S, Daamen W, *Pedestrian Behavior at Bottlenecks*, 2005 TRANSPORTATION SCIENCE **39** 147-159
[2] Seyfried A, Passon O, Steffen B, Boltes M, Rupprecht T and Klingsch W, *New Insights into Pedestrian Flow Through Bottlenecks*, 2009 Transportation Science **43** 395-406
[3] Zhang J, Klingsch W, Schadschneider A and Seyfried A, *Ordering in bidirectional pedestrian flows and its influence on the fundamental diagram*, 2012 Journal of Statistical Mechanics: Theory and Experiment P02002
[4] Guo N, Ding J, Ling X, Shi Q and Takashi I, *The walking behavior of pedestrian crowd under impact of static and movable targets*, 2013 The European Physical Journal B **86** 310
[5] Lian L P, Mai X, Song W G, Richard Y, Wei X G and Ma J, *An experimental study on four-directional intersecting pedestrian flows*, 2015 Journal of Statistical Mechanics: Theory and Experiment P08024
[6] Pastor J M, Garcimartín A, Gago P A, Peralta J P, Martín-Gómez C, Ferrer L M,... Zuriguel I, *Experimental proof of faster-is-slower in systems of frictional particles flowing through constrictions*,

2015 *Physical review. E, Statistical, nonlinear, and soft matter physics* **92** 062817

[7] Garcimartín A, Pastor J M, Martín-Gómez C, Parisi D and Zuriguel I, *Pedestrian collective motion in competitive room evacuation*, 2017 *Scientific Reports* **7** 10792

[8] Steffen B and Seyfried A, *Methods for measuring pedestrian density, flow, speed and direction with minimal scatter*, 2010 *Physica A: Statistical Mechanics and its Applications* **389** 1902-1910

[9] Seyfried A, Passon O, Steffen B, Boltes M, Rupprecht T and Klingsch W, *New Insights into Pedestrian Flow Through Bottlenecks*, 2009 *Transportation Science* **43** 395-406

[10] Zhang J, Kligsch W, Schadschneider A and Seyfried A, *Transitions in pedestrian fundamental diagrams of straight corridors and T-junctions*, 2011 *Journal of Statistical Mechanics: Theory and Experiment* P06004

[11] Kretz T, Grünebohm A and Schreckenberg M, *Experimental study of pedestrian flow through a bottleneck*, 2006 *Journal of Statistical Mechanics: Theory and Experiment* P10014

[12] Gwynne S M V, Kuligowski E D, Kratchman J and Milke J A, *Questioning the linear relationship between doorway width and achievable flow rate*, 2009 *Fire Safety Journal* **44** 80-87

[13] Wagoum A U K and Seyfried A, *Conception, Development, Installation and Evaluation of a Real Time Evacuation Assistant for Complex Buildings*, 2013 *Procedia - Social and Behavioral Sciences* **104** 728-736

[14] Portz A and Seyfried A, *Modeling Stop-and-Go Waves in Pedestrian Dynamics*, 2009 *Parallel Processing and Applied Mathematics* 561-568

[15] Zhao Y, *Verification and validation of the evacuation model[D]*, 2018

[16] Kuligowski E, Peacock R, Wiess E and Hoskins B, *Stair evacuation of older adults and people with mobility impairments*, 2013 *Fire Safety Journal* **62** 230-237

[17] Lachapelle U and Cloutier M, *On the complexity of finishing a crossing on time: Elderly pedestrians, timing and cycling infrastructure*, 2017 *Transportation Research Part A: Policy and Practice* **96** 54-63

[18] Oxley J A, Ihsen E, Fildes B N, Charlton J L and Day R H, *Crossing roads safely: An experimental study of age differences in gap selection by pedestrians*, 2005 *Accident Analysis & Prevention* **37** 962-971

[19] Duim E, Lebrão M L and Antunes J L F, *Walking speed of older people and pedestrian crossing time*, 2017 *Journal of Transport & Health* **5** 70-76

[20] Coffin A and Morrall J, *Walking Speeds of Elderly Pedestrians at Crosswalks*, 1995 *Transportation Research Record,* **1487** 63-67

[21] Gorrini A, Vizzari G and Bandini S, *Age and Group-driven Pedestrian Behaviour: from Observations to Simulations*, 2016 *Collective Dynamics* **1** A3:1-16

[22] Zhang J, Cao S, Salden D and Ma J, *Homogeneity and Activeness of Crowd on Aged Pedestrian Dynamics*, 2016 *Procedia Computer Science* **83** 361-368

[23] Cao S, Zhang J, Salden D, Ma J, Shi C and Zhang R, *Pedestrian dynamics in single-file movement of crowd with different age compositions*, 2016 *Phys Rev E* **94** 012312

[24] Cao S C, Zhang J, Song W G, Shi C A and Zhang R F, *The stepping behavior analysis of pedestrians from different age groups via a single-file experiment*, 2018 *Journal of Statistical Mechanics: Theory and Experiment* P033402

[25] Porzycki J, Mycek M, Lubaś R and Wąs J, *Pedestrian Spatial Self-organization According to its Nearest Neighbor Position*, 2014 *Transportation Research Procedia* **2** 201-206

[26] Webb J D and Weber M J, *Influence of Sensory Abilities on the Interpersonal Distance of the Elderly*, 2016 *Environment and Behavior* **35** 695-711

[27] Winogrond I R, *A Comparison of Interpersonal Distancing Behavior in Young and Elderly Adults*, 1995 *The International Journal of Aging and Human Development* **13** 53-60

[28] Galiza R J, Ferreira L and Charles P, *Estimating the effects of older people in pedestrian flow: A micro- simulation approach*, 2011 *Transport Research Board TRB 90th Annual Meeting. Transportation Research Board (TRB)* 1-13

[29] Spearpoint M and MacLennan H A, *The effect of an ageing and less fit population on the ability of people to egress buildings*, 2012 *Safety Science* **50** 1675-1684

[30] Liu C, Ye R, Lian L P, Song W G, Zhang J and Lo S M, *A least-effort principle based model for heterogeneous pedestrian flow considering overtaking behavior*, 2018 *Physics Letters A* **382** 1324-1334

[31] Shimura K, Ohtsuka K, Vizzari G, Nishinari K and Bandini S, *Mobility analysis of the aged pedestrians by experiment and simulation*, 2014 *Pattern Recognition Letters* **44** 58-63

[32] Boltes M, Sefried A, Steffen B and Schadschneider A, *Automatic Extraction of Pedestrian Trajectories from Video Recordings*, 2010 *Pedestrian and Evacuation Dynamics 2008* 43-54

[33] Hankin B D and Wright R A, *Passenger flow in subways*, 1958 *Oper. Res. Q.* **9** 81

[34] Zhang J, Klingsch W, Schadschneider A and Seyfried A, *Transitions in pedestrian fundamental diagrams of straight corridors and T-junctions*, 2011 *Journal of Statistical Mechanics: Theory and Experiment* P06004